# New determination of the Nuclear Quadrupole Moments of In isotopes

L. A. Errico[1], M. Weissmann[2], and M. Rentería[1]

[1] Departamento de Física, Facultad de Ciencias Exactas,
Universidad Nacional de La Plata, CC No 67, 1900 La Plata, Argentina.
[2] Departamento de Física, Comisión Nacional de Energía Atómica,
Avda. del Libertador 8250, 1429 Buenos Aires, Argentina.

We present here a new determination of the quadrupole moment $Q$ of hyperfine-probe-nuclei states of three different In isotopes: the $5^+$ 192 keV excited state of $^{114}$In (probe for Nuclear Orientation experiments), the $9/2^+$ ground state of $^{115}$In (Nuclear Magnetic and Nuclear Quadrupole Resonance probe) and the $3/2^+$ 659 keV excited state of $^{117}$In (Perturbed-Angular-Correlations probe). These nuclear quadrupole values were determined by comparing experimental nuclear-quadrupole frequencies with the electric-field-gradient tensor (EFG) calculated at In sites in metallic Indium within the Density-Functional Theory. These accurate *ab initio* calculations were performed with the WIEN97.10 implemenattion of the Full-Potential Linearized-Augmented Plane-Wave method. The obtained results for the quadrupole moments [$Q(^{114}\text{In}) = -0.14(1)$ b; $Q(^{115}\text{In}) = -0.77(1)$ b; $Q(^{117}\text{In}) = 0.59(1)$ b] are in clear discrepancy with those reported in the literature [$Q(^{114}\text{In}) = +0.14(6)$ b; $Q(^{115}\text{In}) = +0.861(45)$ b; $Q(^{117}\text{In}) = 0.64(4)$ b].

**Introduction**

The experimental study of nuclear-quadrupole interactions is often used as a powerful tool for characterizing different atomic sites in a given sample and to obtain information about local symmetry, coordination and valence of defect or structural centers in solids, among other electronic, structural or magnetic properties [1]. In the case of pure electric-quadrupole interactions, the measured quantities are the quadrupole coupling constant $\nu_Q = eQV_{33}/h$ and the asymmetry parameter $\eta = (V_{11}-V_{22})/V_{33}$, where $V_{ii}$ ($i = 1, 2, 3$) are the components of the diagonalized electric-field gradient (EFG) tensor with the convention $|V_{33}|>|V_{22}|>|V_{11}|$. In this way, $V_{33}$ is the mayor principal component of the EFG tensor. The EFG is measured via its interaction with the nuclear-quadrupole moment $Q$ of a suitable probe-atom (generally an impurity in the system under study) by different techniques, such as Mössbauer Spectroscopy, nuclear magnetic and Nuclear Quadrupole Resonance (NMR and NQR) or Perturbed-Angular Correlations (PAC) [2-4]. Since the EFG tensor is directly related to the asphericity of the electronic density in the vicinity of the probe nucleus, the quadrupole coupling constant $\nu_Q$ allows the estimation of covalency or ionicity of chemical bonds in the solid, provided $Q$ is known. Although $Q$ is a purely nuclear quantitity, for some isotopes the quadrupole moments are know only with limited accuracy and their determination is still an active field of research.

One of the possible methods to determine $Q$ is to use information of both experimental quadrupole coupling constant and reliable theoretical EFG calculations for the same system. These calculations are very difficult, because the EFG tensor depends strongly on the occupation of orbitals with different symmetry around the nucleus. In effect, small differences in the order of 0.01 electrons in the occupation of a given orbital led to quite different EFG results. For this reason the EFG calculations require a very good description of the electronic structure of the system under study, which depending on the system being considered should be not a simple task.

For a long time the EFG have been calculated using the simple point-charge model (PCM). Since such calculations do not account for any onsite polarization, atomic Sternheimer antishielding factors are introduced to describe core polarization, charge transfer effects are only crudely estimated, and covalence is completely neglected. In their pioneering work, Blaha, Schwarz, and Herzig [5] showed that the Linearized-Augmented Plane-Wave band-structure method (LAPW) [6] was able to predict with high accuracy the EFG in simple solids without the use of external factors, like the Sternheimer one. In the last decade, increasing computer power and progress in methodological development enable calculations in more complicated systems, for example ionic insulators like $Li_3N$ [6], semiconductors oxides like $Cu_2O$ [7] or $TiO_2$ [8], hcp metals [9], high $T_c$ superconductors [10], or to model surfaces [11] or impurity systems [12, 13]. Moreover, this method proved to be both accurate and reliable to determine nuclear-quadrupole moments of even quite heavy nuclei by comparison of EFG calculations with the experimentally measured $n_Q$ [14].

In this work we present a new determination of the nuclear-quadrupole moments of three nuclear states of In isotopes: the $5^+$ 192 keV excited state of $^{114}$In, the $9/2^+$ ground state of $^{115}$In and the $3/2^+$ 659 keV excited state of $^{117}$In. All these nuclear states are used as nuclear probes in Nuclear Orientation (NO) experiments ($^{114}$In), Nuclear Magnetic (NMR) and Nuclear Quadrupole Resonance (NQR) ($^{115}$In), and Perturbed-Angular-Correlations (PAC) experiments ($^{117}$In). These determinations were obtained by comparing experimentally determined nuclear quadrupole frequencies with the EFG calculated at In sites in metallic indium using the Full-Potential LAPW method (FLAPW). As we will show, the obtained results are in clear discrepancy with those reported in the literature.

**Method and computational aspects.**

The EFG tensor is a traceless symmetric tensor of rank two, defined as the second derivative (with respect to the spatial coordinates) of the Coulomb potential at the position of the nucleus. The Coulomb potential can be determined from the total charge density in the crystal by solving Poisson's equation. In this scheme, the EFG can be determined straightforward once the total charge distribution has been calculated. For instance, the principal component $V_{33}$ of the diagonalized EFG tensor at a probe-nucleus located at the origin is given by:

$$V_{ZZ} = \int \rho(\bar{r}) \frac{2 P_2 (\cos(\theta))}{r^3} dr , \qquad (1)$$

where $P_2$ is the second-order Legendre polynomial and $\rho(r)$ is the total charge density. As can be seen from the above equation, the charge density is weighted by a factor $1/r^3$ and thus, the theoretical method employed must be able to describe accurately the wavefunctions in the immediate vicinity of the nucleus (at sub-nanoscopic scale) in which the EFG is calculated. For this reason, pseudopotential methods are not well suited for EFG calculations. So, a cluster method (with a sufficiently large cluster size to ensure the accurate description of the total charge density in the vicinity of the nucleus) or band-structure methods are the correct way to calculate the EFG tensor. A More detailed description of the EFG calculation within the FLAPW method can be found in Ref. 15.

For the calculations presented in this paper we employed the full-potential linearized-augmented plane-wave (FP-LAPW) method as embodied in the WIEN97.10 code [16] in a scalar relativistic version without spin-orbit coupling. In this method, no shape approximation on either the potential or the electronic charge density is made, being thus specially suited for EFG calculations. For methodological purposes the unit cell is divided into non-overlapping spheres with radius $R_i$ and an interstitial region. In the later the wavefunctions are expanded into plane-waves, while inside the spheres these plane-waves are augmented by an atomic-like spherical harmonics expansion. The atomic spheres radii used for In was 1.3 Å An advantage of plane-wave-based methods is that the convergence of their basis-set can easily be tested by including additional plane waves in the calculations. This was done for several cases here and well-converged solutions were found when the parameter $RK_{MAX}$ (which controls the size of the basis-set in these calculations) was equal to 9. This value gives well converged basis-set consisting of 150 LAPW functions. In addition to the usual LAPW basis-set, we also introduced local orbitals (LO) to include In-4$d$ orbitals. Integration in reciprocal space were performed using the tetrahedron method taking 30,000 $k$-points in the full Brillouin zone (BZ), which are reduced to 2176 $k$-points in the irreducible wedge of the BZ. The correctness of the choice of these parameters was checked by performing calculations for other $R_i$, $k$-point sampling, and $RK_{MAX}$ values. In the next section we will give a more detailed description of the convergence studies. Finally, exchange and correlation effects were treated within Density-Functional Theory using either the local-density approximation (LDA) [17] or the generalized-gradient approximation (GGA) [18].

**Results for the EFG tensor and determination of the nuclear-quadrupole moments.**

Metallic Indium crystallizes in a distorted cubic close packing. The unit cell is tetragonal, $a = b = 3.221$ , $c = 4.933$ at 4.2 K ($a = b = 3.253$ , $c = 4.947$ at 300 K) [19], with two atoms in the body-centered positions (0 0 0) and (1/2 1/2 1/2). In this metal, due to the mentioned distortion, the 12 equidistant neighbors of the cubic close-packing fall into one group of four at a distance of 3.24 and a group of eight not so near, at 3.37 .

Initially, we calculated the equilibrium lattice parameters predicted by FP-LAPW. To obtain the equilibrium lattice parameters we mapped the energy surface as function of $a$ and $c$ in order to obtain the lattice parameters corresponding to the minimum energy of the system. This minimum correspond to $a = 3.178$ and $c = 4.887$ , in excellent agreement with the experimental determinations, showing that FLAPW describes correctly the structure of metallic Indium, given support to our EFG calculations.

Now we can present our results for the EFG tensor. We obtain (for the experimental lattice parameters at 4.2 K and for the muffin tin radius, $k$-point sampling, and $RK_{MAX}$ values detailed in the previous section) $V_{33} = +2.45 \times 10^{21}$ V/m$^2$ (for LDA) and $V_{33} = +2.39 \times 10^{21}$ V/m$^2$ (GGA). In both cases, and as it was expected according to the symmetry of the structure, we found $\boldsymbol{h} = 0.0$, in perfect agreement with the experimental results. The orientation of $V_{33}$ is parallel to the $c$-axis, meanwhile $V_{22}$ are parallel to the $a$ axis (of course, due to the EFG tensor symmetry, $V_{11}$ points parallel to the equivalent $b$ axis).

In order to evaluate the convergence of our results, we performed calculations for different muffin tin radii, $k$-point sampling and basis set size. The muffin-tin radius was varied from 1.1 to 1.6 ., meanwhile, in the case of the $k$-point sampling, we performed calculations increasing the sampling from 10,000 $k$-points in the BZ (726 $k$-points in the irreducible wedge of the BZ) to 45,000 $k$-points (3,078 $k$-points in the irreducible wedge of the BZ). In order to check the convergence in the basis set we performed calculations from $RK_{MAX} = 6$ (45 LAPW functions for $Ri = 1.3$ ) to $RK_{MAX} = 10$

(200 LAPW functions for the same radius). The conclusion of all these calculations is that, for the $Ri = 1.3$, 30,000 $k$-points and $RK_{MAX} = 9$, the convergence error in the EFG tensor components are smaller than $0.05 \times 10^{21}$ V/m$^2$. We will take this value as the convergence error in our calculation.

Once we presented the results for the EFG tensor, we can present our results for the nuclear quadrupole moments of In isotopes.

a) The 5$^+$ 192 keV excited state of $^{114}$In: The nuclear quadrupole interaction of In in metallic Indium was determined by Brewer *et al.* [20] using the Nuclear Quadrupole Alignment (NQA) technique. While the sign determination is lacking in most experimental techniques, NQA gives a direct and straightforward determination of the sign of the quadrupole frequency $n_Q$. The obtained result (at 4.2 K) was $n_Q$ = -8.4(3) MHz. As it was expected, the measured value of the asymmetry parameter was $h$ = 0.0. Since the nuclear quadrupole moment of the 192 keV excited state was not determined when this experiment was performed, from the above frequency the authors obtained a value of $Q(^{114}$In$)$ = +0.16(6) b, assuming the EFG to be negative [20]. Unfortunately, the authors do not discuss the method of calculation employed for the calculation of the EFG tensor (probably, point-charge calculations in combination with corrective factors).

If we use our FP-LAPW results for the EFG tensor at In sites in metallic Indium and the reported experimental values of $n_Q$, we can obtain from $n_Q = eQV_{33}/h$ a value of $Q$ = -0.14(1) b in the case of LDA and $Q$ = -0.15(1) b in the case of GGA, in clear contradiction with the previously reported value [20].

b) The 9/2$^+$ ground state of $^{115}$In: The magnitude of $n_Q$ in metallic In has been accurately determined over a large temperature range using NQR and the ground state of $^{115}$In as probe. The obtained result was $|n_Q|$ = 45.24(1) MHz at 4.2 K [19, 21, 22]. The magnitude of $n_Q$ was also determined at 300 K using NMR. The obtained result was $|n_Q|$ = 29.50 MHz [23] (reported without error). Magnitude and sign of $n_Q$ were determined by F. C. Thatcher and R. R. Hewitt. The reported result is $n_Q$ = -45.36 MHz (reported without error) [24]. Using this experimental value and our FLAPW results for the EFG, we obtain $Q$ = -0.76(2) b in the case of LDA and $Q$ = -0.78(2) b (GGA).

The nuclear quadruple moment of the ground state of $^{115}$In has been obtained by G. F. Koster [25], who studied the effects on the hyperfine structure of the mixing of higher configurations with the ground state of $^{115}$In. By combination of this calculation and atomic beam technique results, a value of $Q$ = +0.834 b was obtained. As in the case of $^{114}$In, this result for the nuclear quadrupole moment of the ground state of $^{115}$In is in clear contradiction with those obtained in the present work.

c) The 3/2$^+$ 659 keV excited state of $^{117}$In: The 659 keV excited state of $^{117}$In can be use as sensitive state in $g$-$g$ PAC experiments. Using this technique, the magnitude of $n_Q$ in metallic In has been determined at different temperatures. The result obtained by R. S. Raghavan and P. Raghavan at 4.2 K is $n_Q$ = 32.1(5) MHz, $h$ = 0.0 ($n_Q$ = 21.74(22) MHz at 295 K) [26]. Due to the characteristics of the $g$-$g$ PAC technique, only the magnitude and symmetry of the quadrupole frequency has been determined. Hence it will not be possible to determined the sign of Q for this measurement.

As we said before, the usual problem in the determination of the quadrupole moments is the lack of reliable estimations of the EFG tensor although the quadrupole frequency interaction could be measured accurately. R. S. Raghavan and P. Raghavan eliminated the need to know the EFG by the fact that NQR resonance experimental results on $^{115}$In in In$_{0.99}$Cd$_{0.01}$ exist [26]. In effect, Thatcher and Hewitt [24] have made NQR measurements in this compound at 4.2 K. They report $n_Q(^{115}$In$)$ = 43.2(1) MHz. Thus, the ratio $|Q(^{117}$In$)/Q(^{115}$In$)|$ is derived to be 0.743(15). Using the value of the nuclear quadrupole moment of the ground state of $^{115}$In reported by G.F. Koster [25] (after applying a Sternheimer correction of 3.2%), R. S. Raghavan and P. Raghavan obtained $|Q(^{117}\text{In})| = 0.64(4)$ b [26]. A similar result was obtained by H. Haas and D. A. Shyrley [27]. Using the experimental value of $n_Q$ at 4.2 K and our first-principles theoretical values for $V_{33}$, we found $|Q| = 0.58(1)$ b (LDA), $|Q| = 0.60(1)$ b (GGA), without the use of arbitrary corrections.

**Conclusions**

In the present work we have presented *ab initio* FP-LAPW results for the EFG tensor at In sites in metallic Indium. The combination of experimental hyperfine interaction measurements with this theoretical EFG calculations enable us to determine the nuclear quadrupole moment of the different states of In isotopes, that can be used as sensitive states in different experimental techniques. Our results are in clear contradiction with those reported in the literature. We think that these results should stimulate new investigations of the *Q* values use in the field of hyperfine techniques.

**Acknowledgements**

This work was partially supported by CONICET, Fundación Antorchas, and ANPCyT (PICT98 03-03727), Argentina, and TWAS, Italy. L. A. E. is fellow of CONICET. M.W. and M.R. are members of CONICET.

**References**

[1] See, e.g,, Proceedings of the 12th International Conference on Hyperfine Interactions, Park City, Utah, 2001, edited by W.E. Evenson, H. Jaeger, and M.O. Zacate, Hyp. Interactions **136/137**, 2001.
[2] G. Schatz and A. Weidinger, in *Nuclear Condensed Matter Physics - Nuclear Methods and Applications*, edited by John Wiley \& Sons (Chichester, England, 1996) p.63.
[3] E.N. Kaufmann and R.J. Vianden, *Rev. Mod. Phys.* **51**, 161 (1979).
[4] H. Frauenfelder and R. Steffen, in *a-, b-, and g-Ray Spectroscopy*, edited by K. Siegbahn (North-Holland, Amsterdam, 1968), Vol. 2, p. 917.
[5] P. Blaha, K. Schwarz, and P. Herzig, Phys. Rev. Lett. **54**, 1192 (1985).
[6] S.H. Wei and H. Krakauer, Phys. Rev. Lett. **55**, 1200 (1985).
[7] K. Schwarz and P. Blaha, Z. Naturforsch. A **47**, 197 (1992).
[8] P. Blaha, D. J. Singh, P. I. Sorantin, and K. Schwarz, Phys. Rev. B **46**, 1321 (1992).
[9] P. Blaha, K. Schwarz, and P. H. Dederichs, Phys. Rev. B **37**, 2792 (1988).
[10] K. Schwarz, C. Ambrosch-Draxl, and P. Blaha, Phys. Rev. B **42**, 2051 (1990).
[11] C.O. Rodriguez, M. V. Ganduglia-Pirovano, E.L. Peltzer y Blancá, and M. Petersen, Phys. Rev. B **64**, 144419 (2001).
[12] L.A. Errico, G. Fabricius, M. Rentería, P. de la Presa, and M. Forker, Phys. Rev. Lett. **89**, 55503 (2002).
[13] L.A. Errico, G. Fabricius, and M. Rentería, Phys. Rev. B **67**, 144104 (2003) and references therein.

[14] P. Dufek, P. Blaha, and K. Schwarz, Phys. Rev. Lett. **75**, 3545 (1995); P. Blaha, P. Dufek, K. Schwarz, and H. Haas, Hyp. Interactions **97/98**, 3 (1996); P. Blaha, K. Schwarz, W. Faber, and J. Luitz, Hyp. Interactions **126**, 389 (2000).
[15] K. Schwarz, C. Ambrosch-Draxl, and P. Blaha, Phys. Rev. B **42**, 2051 (1990).
[16] P. Blaha, K. Schwarz, P. Dufek, and J. Luitz, *wien97*, Vienna University of Technology, 1997. Improved and updated Unix version of the original copyrighted *Wien*-code, which was published by P. Blaha, K. Schwarz, P.I. Sorantin, and S. B. Trickey, in Comput. Phys. Commun. **59**, 399 (1990).
[17] J. P. Perdew and Y. Wang, Phys. Rev. B **45**, 13244 (1992).
[18] J.P. Perdew, K. Burke, and M. Ernzerhof, Phys. Rev. Lett. **77**, 3865 (1996).
[19] R. R. Hewitt and T. T. Taylor, Phys. Rev. **125**, 524 (1962).
[20] W. D. Brewer and G. Kaindl, Hyp. Interactions **4**, 576 (1978);
[21] W. W. Simmons and C. P. Slichter, Phys. Rev. **121**, 1580 (1961).
[22] W. J. O'Sullivan and J. E. Schirber, Phys. Rev. **135**, A1261 (1964).
[23] D. R. Torgeson and R. G. Barnes, Phys. Rev. Lett. **9**, 255 (1962).
[24] F. C. Thatcher and R. R. Hewitt, Phys. Rev. B. **1**, 454 (1970).
[25] G. F. Koster, Phys. Rev. **86**, 148 (1952).
[26] R. S. Raghavan and P. Raghavan, Phys. Rev. Lett. **28**, 54 (1972).
[27] H. Haas and D. A. Shirley, The Journal of Chemical Physics **58**, 3339 (1973).